\documentstyle[12pt,epsf]{article}

\title{Long-range acceleration induced by a scalar field external to gravity and
the indication from Pioneer 10/11, Galileo and Ulysses Data}

\author{J.P.  Mbelek and M.  Lachi\`eze-Rey \\ Service d'Astrophysique, C.E.
Saclay \\ F-91191 Gif-sur-Yvette Cedex, France}

\begin{document} \maketitle \baselineskip=8mm

\begin{abstract} The anomalous acceleration $a_{P}$ of the Pioneer 10/11 - the
Pioneer effect - has remained unexplained.  We suggest an explanation based on
the interaction of the spacecraft with a long-range scalar field, $\phi$.  The
scalar field under consideration is external to gravity, coupled to the ordinary
matter and undergoes obedience to the equivalence principle.  In addition to its
self-interaction term, the field is determined by an external source term
proportional to the Newtonian potential in the weak fields limit it result a
long-range acceleration $a_{P}$, asymptotically constant within the region of
the solar system hitherto crossed by the spacecraft.  Also, the limit
$0.1~10^{-8}$ cm/s$^{2}$, which follows from the Viking ranging data, is
satisfied in the region of the terrestrial planets (in particular at the
positions of the Earth and Mars), for a $\phi$-field of mass $m_{\phi} \geq
1.8~10^{-17}~eV/c^{2}$.  The proposed solution gives the correct order of
magnitude for $a_{P}$, as observed so far, and predicts the decline of $a_{P}$
in the form of damped oscillations beyond $97$ AU.  An estimate of the
cosmological constant is also made by taking into account the contribution of
the vacuum energy density of the scalar field in galactic dynamics and
particularly in the outskirts of the dwarf galaxy DDO 154.  \end{abstract}

\section{Introduction}\label{intro} Recently, results from an almost twenty
years study of radio metric data from Pioneer 10/11, Galileo and Ulysses
spacecraft have been published by a team of the NASA (Anderson et
al.~\cite{Andersona}), indicating an apparent anomalous, constant, acceleration
acting on the spacecraft with a magnitude of the order $8.5~10^{-8}$ cm/s$^{2}$,
directed towards the Sun, to within the accuracy of the Pioneers' antennas and a
steady frequency drift, a "clock acceleration", of about $-6~10^{-9}$ Hz/s.  A
number of potential causes have been ruled out by the authors, namely gravity
from Kuiper belt, gravity from the Galaxy, spacecraft "gas leaks", anisotropic
heat (coming from the RTGs) reflection off of the back of the spacecraft
high-gain antennae (Katz's proposal~\cite{Katz}, see Anderson et
al.~\cite{Andersonb}), radiation of the power of the main-bus electrical systems
from the rear of the craft (Murphy's proposal~\cite{Murphy}, see Anderson et
al.~\cite{Andersonc}), errors in the planetary ephemeris, and errors in the
accepted values of the Earth's orientation, precession, and nutation, as well as
nongravitational effects such as solar radiation pressure, precessional
attitude-control maneuvers and a possible nonisotropic thermal radiation due to
the Pu$^{238}$ radioactive thermal generators.  Indeed, according to the
authors, none of these effects explain the apparent acceleration and some are 3
orders of magnitude or more too small, so they conclude that there is an
unmodeled acceleration towards the Sun of $(8.09 \pm 0.20) 10^{-8}$ cm/s$^{2}$
for Pioneer 10, $(8.56 \pm 0.15) 10^{-8}$ cm/s$^{2}$ for Pioneer 11, $(12 \pm 3)
10^{-8}$ cm/s$^{2}$ for Ulysses and $(8 \pm 3) 10^{-8}$ cm/s$^{2}$ for Galileo.
The authors plan to utilize two different transmission frequencies in further
analysis to give an answer to whether there is some unknown interaction of the
radio signals with the solar wind.

Since no "standard physics" plausible explanations for the residual acceleration
has been found so far, the authors considered the possibility that the origin of
the anomalous signal is the effect of a modification of gravity, for instance by
adding a Yukawa force to the Newtonian or Milgrom's proposed modification of
gravity (Milgrom~\cite{Milgrom}).  They concluded however that neither easily
works.

If the cause is dark matter, the amount needed to be consistent with the
accuracy of the ephemeris should be only of order a few times $10^{-6}
M_{\odot}$ even within the orbit of Uranus (Anderson et al.~\cite{Andersond}).
Above all, the authors point out that the residual acceleration is too large to
have remained undetected in the planetary orbits of the Earth and Mars.  Indeed,
the Viking ranging data limit any unmodeled radial acceleration acting on the
Earth and Mars to no more than $0.1~10^{-8}$ cm/s$^2$.  Because of this severe
constraint, the authors argue that, if the anomalous radial acceleration is of
gravitational origin, it probably violates the principle of equivalence.  But,
an alternative is that the anomalous acceleration is asymptotically constant,
rather than constant at all radii from the center of the solar system to the
present location of Pioneer 10/11 spacecraft.

In this paper we propose an alternative explanation, based on the possible
existence of a long range (non gravitational) scalar field, $\phi$, which
respects the (weak) equivalence principle.  This possibility was previously
introduced by Mbelek~\cite{Mbelek}, to account for the rotational curves of
spiral galaxies, as an alternative for dark matter.  It gives, for the Pioneer
10/11 spacecraft, the correct order of magnitude for both the anomalous
acceleration, $a_{P}$, and the clock acceleration, $a_{t}$.  It proves to remain
consistent with the planetary orbits determined from the Viking data.

As for the ordinary matter, the $\phi$-field is a gravitational source through
its energy-momentum tensor.  A forthcoming paper will present the fundamental
symmetry that may support it, from the background of classical fields theory.
The plan of this paper is as follows :  in section 2, we set the $\phi$-field
equation.  Then, after linearization we divide space in three characteristic
regions and find approximate exterior solutions for a static spherically
symmetric source, actually the Sun.  In section 3, Einstein equations are solved
in the weak fields approximation to account for the metric tensor in the
presence of the $\phi$-field (out of the Sun).  In section 4, the equation of
motion of a test body in the presence of the $\phi$-field is established.
Solutions are found in the weak fields and low velocity limit.  Then the
anomalous long-range acceleration $a_{P}$ is derived for the different regions
of space.  In section 5, an interpretation of the data is proposed.  In section
6, the steady frequency drift $a_{t}$ is derived by using the equivalence
principle.  We finally conclude by an estimation of the cosmological constant by
exploiting the declining part of the rotational curve (RC) of the dwarf galaxy
DDO 154.

\section{The scalar field equation}

A manner to generate a constant radial acceleration could result from the
introducion of a linear potential term in the Lagrangian of a test particle.  An
example is provided by the exterior solution of the locally conformal invariant
Weyl gravity for a static, spherically symmetric source (Mannheim and
Kazanas~\cite{Mannheim}).  Unfortunately, Perlick and Xu~\cite{Perlick}, by
matching the exterior solution to an interior one that satisfies the weak energy
condition and a regularity condition at the center, show that this leads to
contradiction of Mannheim and Kazanas's suggestion.  They conclude that the
conformal Weyl gravity is not able to give a viable model of the solar system.
\\ This paper presents an alternative solution, under the form of a real scalar
field, external to gravity but which satisfies the equivalence principle.  We
show below that it leads to the desired " Pioneer effect ", although it does not
modify, as required, the orbital properties of the internal planets.  The field
$\phi$ obeys the equation \begin{equation} \label{scalar field GR eq}
{\nabla}_{\nu} {\nabla}^{\nu} {\phi} = - U'(\phi) - J,\end{equation} where the
symbol ${\nabla}_{\nu}$ stands for the covariant derivative compatible with the
Levi-Civita connection.  Equation (\ref{scalar field GR eq}) may be derived from
Einstein equations provided that the energy-momentum tensor of the $\phi$-field
is of the form $T^{(\phi)}_{\mu\nu} = {{\partial}_{\mu}}{\phi}
~{{\partial}_{\nu}}{\phi} - g_{\mu\nu} [\frac{1}{2} {{\partial}_{\lambda}}{\phi}
~{{\partial}^{\lambda}}{\phi} - U(\phi) - \int J d\phi]$ (up to a positive
multiplicative dimensionality constant, $\kappa$, for $\phi$ is dimensionless in
this paper) and the energy-momentum tensor of the ordinary matter (matter or
radiation other than the $\phi$-field) is divergenceless (e.g., zero for the
exterior solution and of the perfect fluid form for the interior solution).\\In
the rest of the paper, we apply the weak field approximation to the real
classical scalar field $\phi$ and to the gravitational potentials, so that
Newtonian physics apply.  For a weak gravitational field, equ(\ref{scalar field
GR eq}) above will write merely \begin{equation} \label{scalar field SR eq}
{\partial}_{\mu} {\partial}^{\mu} {\phi} = - U'(\phi) - J.\end{equation} The
potential $U$ denotes the self-interaction of $\phi$, and we note $U'(\phi) =
\frac{\partial U} {\partial {\phi}}$.  The source term $J$, an external source
function, takes gravity into account, as a source for the field $\phi$.  Of
course, $\phi$ also acts as a source for gravity (through Einstein equations).
We consider this latter action below in section $3$.

In the weak field approximation, $J$ depends on the Newtonian gravitational
potential $V_{N}$, the only relevant scalar quantity related to a weak
gravitational field.  Thus, we write at first order $J = J(\frac{V_{N}}{c^{2}})
\approx - \frac{V_{N}}{r_{0}^{2}~c^{2}}$, where the constant $r_{0}$ defines a
characteristic length scale (see subsection $5.1.1$ for an estimation of
$r_{0}$).  The minus sign comes from the requirement that the effect of $\phi$
is similar to that of gravitation, so that $\frac{d\phi}{dr}$ and
$\frac{dV_{N}}{dr}$ have the same sign (as we will see, the $\phi$-field
generates an acceleration term $\frac{d\phi}{dr}$, up to a positive
multiplicative factor).  This is in accordance with our previous study on the RC
of spiral galaxies in which we found the positivity of $\frac{d\phi}{dr}$
necessary for the $\phi$-field mimics a great part of the missing
mass~\cite{Mbelek}.  In the solar system, $\frac{d\phi}{dr}$ remains positive.
The scalar field $\phi$ is positive definite throughout this paper.

Here we explore the effect of $\phi$ in the solar system, i.  e., in the
potential $V_{N} = - c^{2}~r_{s}/2r$ created by the static central mass of the
the Sun, $r$ being the radius from the centre (we choose as usual a zero value
of the Newtonian potential at infinity), and $r_{s}$ the Schwarzschild radius of
the Sun.  The problem has spherical symmetry, so that equation (\ref{scalar
field SR eq}) yields finally \begin{equation} \label{phi eq with U and VN
explicit} \frac{d^{2}{\phi}}{{dr}^{2}} + \frac{2}{r} \frac{d{\phi}}{dr} =
U'(\phi) + \frac{r_{s}} {2r ~r_{0}^{2}}.  \end{equation} We will calculate the
resulting $\phi$-field and show that it creates an asymptotically constant
acceleration:  we solve the equation with the limiting condition (imposed by the
weak fields approximation) that the field $\phi$ and its derivative
$\frac{d\phi}{dr}$ are bounded for any given region of space.  In addition, the
field $\phi$ must vanish (up to an additive constant) if one sets $M = 0$, since
the central mass $M$ is its source ; the same condition applies to
$\frac{d\phi}{dr}$.

As a first step to resolve equ(\ref{phi eq with U and VN explicit}), let us
neglect for the moment the contribution of the self-interaction.  The solution
is \begin{equation} \label{linear sol} \phi = C + \frac{r_{s}}{4{r_{0}}^{2}} r -
\frac{A}{r} \end{equation} and thence, \begin{equation} \label{grad linear sol}
\frac{d\phi}{dr} = \frac{r_{s}}{4r_{0}^{2}} + \frac{A}{r^{2}}, \end{equation}
where A and C are constants of integration.  The constant of integration $A$ of
the dimension of a length obviously depends on $r_{s}$ since the central mass is
the source of the $\phi$-field.  Accordingly, we may set $A = \zeta r_{s}/2$,
where $\zeta$ is a positive dimensionless constant that we will assume hereafter
of the order unity.  The positivity of $\zeta$ is inferred by the positivity of
the spatial derivative $\frac{d\phi}{dr}$ at any distance from the centre.  Note
that the particular value $\zeta = 1$ involves the identity of the potential
term $- A/r$ with the Newtonian one $V_{N}/c^{2}$.  We will show that, in a
certain radius range and at sufficiently large distances from the centre, this
represents the true solution to equ(\ref{phi eq with U and VN explicit}):  the
radial acceleration induced by the $\phi$-field (neglecting its
self-interaction), proportional to $\frac{d\phi}{dr}$, remains asymptotically
constant.  This will be referred to as the "Pioneer effect" throughout.\\In
order to solve the complete equation, we need to know the form of $U(\phi)$,
although we will see that many results remain independent of this choice.  To
illustrate, we choose here a quartic self-interaction potential $U = U(0) +
\frac{1}{2} {\mu}^{2} {\phi}^{2} + \frac{\sigma}{4} {\phi}^{4}$, where $\sigma <
0$ is the self-coupling coefficient of the scalar field ; $\frac{1}{2} {\mu}^{2}
{\phi}^{2}$ is the "mechanical" mass term with $\mu = \frac{m_{\phi}c}{\hbar}$,
$m_{\phi}$ denoting the mass of the scalar field.  The reason for choosing a
quartic polynomial form is that the corresponding quantum field theory should be
renormalizable (Madore~\cite{Madore}).

This potential presents two extrema~:  one minimum at $\phi_{I} =\phi(r_{I})$,
with $U'({\phi}_{I}) = 0$ and $U''({\phi}_{I}) >0$ ; and one maximum at
$\phi_{III} =\phi(r_{III}) =\mu/\sqrt{\mid \sigma \mid}$, with $U'({\phi}_{III})
= 0$ and $U''({\phi}_{III}) < 0$ ($U''(\phi) = \frac{{\partial}^{2} U}{{\partial
\phi}^{2}}$).  There is also an inflexion point at $\phi _{II} = \phi(r_{II}) =
\mu/\sqrt{3 \mid \sigma \mid}$, with $U'({\phi}_{II}) > 0$ but $U''({\phi}_{II})
= 0$.  Since $\phi$ increases monotonically with respect to $r$, this
corresponds to three regions I, II, III in space with ${r}_{I} < {r}_{II} <
{r}_{III}$.  Moreover, the monotony of $\phi$ together with the relation
${\phi}_{III} = \sqrt{3}~{\phi}_{II}$ that links $\phi(r_{III})$ to
$\phi(r_{II})$ involves (using the solution (\ref{linear sol}), in the first
approximation) :  \begin{equation} \label{radii relation} {r}_{III} \geq
\sqrt{3}~{r}_{II} \end{equation}

Let us call ${\phi}_{0}$, generically, a local extremum of $U(\phi)$ or
$U'(\phi)$.  In the neighbour region of space, $\mid \phi - {\phi}_{0} \mid \ll
1$, equation (\ref{phi eq with U and VN explicit}) may be solved in the weak
field approximation by linearizing the function $U'(\phi)$ about ${\phi}_{0}$.
This yields \begin{equation} \label{phi approx eq with U}
\frac{d^{2}{\phi}}{{dr}^{2}} + \frac{2}{r} \frac{d{\phi}}{dr} - U'({\phi}_{0}) -
U''({\phi}_{0}) (\phi - {\phi}_{0}) = \frac{r_{s}}{{2r_{0}}^{2}r} \end{equation}

\begin{itemize}

\item in the first region of space (region I), ${\phi}_{0} = {\phi}_{I}$

\item in region II, ${\phi}_{0} = {\phi}_{II}$

\item in region III, ${\phi}_{0} = {\phi}_{III}$.

\end{itemize}

 Besides, as for the Higgs mechanism of symmetry breaking which too involves a
scalar field and a quartic self-interaction potential, an analogy can be made
with a well known phenomenon in solid state physics~:  the Meissner effect which
is a phase transition of the second kind, between the superconducting to the
normal state.  Here, the field $\phi$ plays the role of the magnetic flux and
$U''({\phi})$ plays the role of the difference $\Delta T = T_{c} - T$ between
the temperature, $T$, of the solid and its critical temperature, $T_{c}$.\\
\begin{itemize}

\item In region I, equ(\ref{phi approx eq with U}) reads \begin{equation}
\label{equI} \frac{d^{2}{\phi}}{{dr}^{2}} + \frac{2}{r} \frac{d{\phi}}{dr} -
{\mu}^{2}~(\phi - {\phi}_{I}) = \frac{r_{s}} {2r~r_{0}^{2}}, \end{equation} with
solution \begin{equation} \label{sol of phi with U I} \phi = {\phi}_{I} -
\frac{{\bar{\lambda}}^{2}}{2{r_{0}}^{2}} \frac{r_{s}}{r} (1 -
e^{-(r-r_{I})/\bar{\lambda}}) \end{equation} which implies \begin{equation}
\label{grad phi general I} \frac{d{\phi}}{dr} =
\frac{{\bar{\lambda}}^{2}}{2{r_{0}}^{2}} \frac{r_{s}}{r^{2}} [1 - (1 +
\frac{r}{\bar{\lambda}}) e^{-(r-r_{I})/\bar{\lambda}}], \end{equation} where
$\bar{\lambda} = 1/\mu$ characterizes the dynamical range of the $\phi$-field in
region I.  Clearly, it is necessary that $r_{I} = 0$ for the solution (\ref{sol
of phi with U I}) be consistent with the conditions on $\phi$ and
$\frac{d\phi}{dr}$.

Let us notice that, for a sufficiently massive $\phi$-field (${m}_{\phi} \geq 10
\sqrt{2}~\hbar/c r_{\odot}=1.8~10^{-17}~eV/c^{2}$, where $r_{\odot}$ denotes the
radius of the central mass), $\frac{d\phi}{dr}$ is smaller by a factor 1/100
than the same quantity that would be involved by relation (\ref{grad linear
sol}).  This means that the $\phi$-field is expelled out from the central region
I, so that the Pioneer effect is destroyed here.

This situation appears analogous to the Meissner effect where the magnetic flux
is expelled out in the superconducting state ($\Delta T > 0$).  That the
$\phi$-field is expelled from region I ($U''({\phi}) > 0$), grants that the
orbits of the internal planets are not modified.  Its significant action on
matter is restricted to regions II and III.  Figure 1 shows the predicted curve
$y = a_{P}/a^{\infty}_{P}$ versus $x = r/\bar{\lambda}$ for region I.

\begin{figure} \centerline{\epsfxsize=12cm \epsfbox{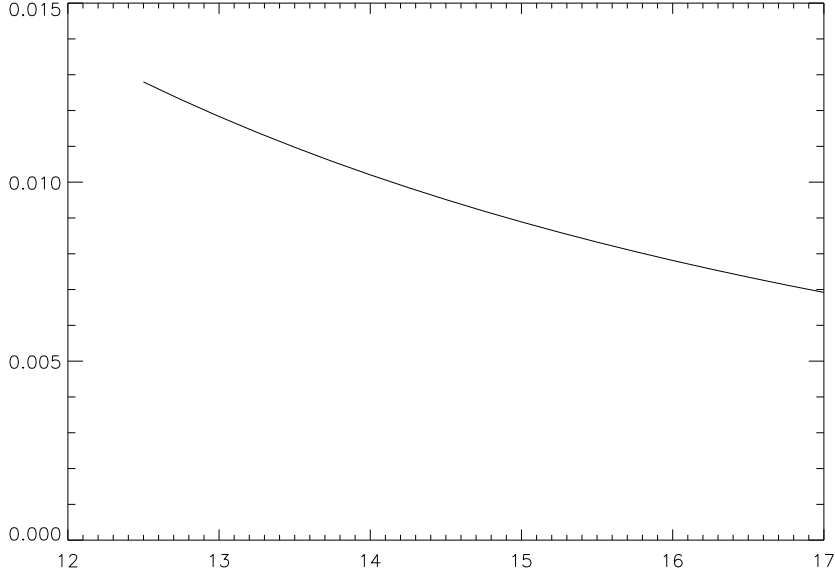}} \caption{Predicted
curve $y = a_{P}/a^{\infty}_{P}$ versus $x = r/\bar{\lambda}$ for region I.  As
one can see, the limiting condition $y \leq 0.01$ is satisfied for $x \geq
10\sqrt{2}$.}  \end{figure}

\item In region II and about $r_{II}$, an approximate solution ${\phi}^{\circ}$
is obtained by solving equation (\ref{phi approx eq with U}) on account that
${\phi}_{0} = {\phi}_{II}$ ; one finds :  \begin{equation} \label{linear sol II}
{\phi}^{\circ} = C + \frac{r_{s}}{4{r_{0}}^{2}} r - \frac{A}{r} +
\frac{U'({\phi}_{II})}{6} r^{2} \end{equation} and \begin{equation}
\label{derivative linear sol II} \frac{d{{\phi}^{\circ}}}{dr} =
\frac{r_{s}}{4{r_{0}}^{2}} + \frac{A}{r^{2}} + \frac{U'({\phi}_{II})}{3} r,
\end{equation} where $A$ and $C$ are constants of integration.  Clearly, the
extra potential $U'({\phi}_{II}) r^{2}/6$ will behave like a positive
cosmological constant type term.  Let us notice that the condition of the weak
field approximation, $\mid \phi - {\phi}_{II} \mid \ll 1$, is always satisfied
as long as $\mid r - r_{II} \mid \ll {r_{0}}^{2}/r_{s}$ in as much as the
$\Lambda$ term is neglected.\\In region II below or beyond $r_{II}$, an improved
solution $\phi = {\phi}^{\circ} + {\delta\phi}$ is obtained in the first
approximation by adding to the previous solution ${\phi}^{\circ}$ a correction
term ${\delta\phi}$.  This involves :  \begin{equation} \label{eq delta phi
I-II-III} \frac{d^{2}{\delta\phi}}{{dr}^{2}} + \frac{2}{r}
\frac{d{\delta\phi}}{dr} - U''({\phi}^{\circ}) \delta\phi = 0
\end{equation}\\Now, the "curvature" $U''(\phi)$, and hence its mean value, is
positive between $r_{I}$ and $r_{II}$ but negative between $r_{II}$ and
$r_{III}$.  In the first approximation, one may write $U''(\phi < {\phi}_{II})
\approx U''({\phi}_{I}) = 1/{\bar{\lambda}}^{2}$ and $U''(\phi > {\phi}_{II})
\approx - k^{2}$, with $k$ a positive constant.  We show in the following that
$\lambda = \frac{2\pi}{k}$ defines a wavelength related to the part of region II
beyond $r_{II}$.  So, replacing $U''({\phi}^{\circ})$ by the value $- k^{2}$,
equation (\ref{eq delta phi I-II-III}) becomes in the first order approximation
:  \begin{equation} \label{eq delta phi II-III mean U''}
\frac{d^{2}{\delta\phi}}{{dr}^{2}} + \frac{2}{r} \frac{d{\delta\phi}}{dr} +
k^{2} \delta\phi = 0.  \end{equation} The solution of the above equation is of
the form \begin{equation} \label{sol delta phi II-III} \delta\phi = \frac{B}{r}
\sin{(kr - {\Phi}_{II})}, \end{equation} where $B$ is a constant of integration
and ${\Phi}_{II}$ is a phase offset.  Consequently, on account of the solution
(\ref{linear sol II}) and the continuity of $\phi$ at the radius $r_{II}$, the
first order solution of equation (\ref{phi eq with U and VN explicit}) writes in
region II beyond $r_{II}$ :  \begin{equation} \label{linear sol general II-III}
{\phi} = C + \frac{r_{s}}{4{r_{0}}^{2}} r - \frac{A}{r} +
\frac{U'({\phi}_{II})}{6} r^{2} + \frac{B}{r} \sin{k(r - r_{II})},
\end{equation} \begin{equation} \label{derivative of phi II-III bis}
\frac{d\phi}{dr} = \frac{r_{s}}{4{r_{0}}^{2}} + \frac{A}{r^{2}} +
\frac{B}{r^{2}} [\frac{2{\pi}r}{\lambda}\cos{(\frac{2{\pi}(r -
r_{II})}{\lambda})} - \sin{(\frac{2{\pi}(r - r_{II})}{\lambda})}] +
\frac{U'({\phi}_{II})}{3} r.  \end{equation} Below $r_{II}$, the solution is of
the form :  \begin{equation} \label{sol delta phi I-II} \phi = {\phi}_{I} -
\frac{{\bar{\lambda}}^{2}}{2{r_{0}}^{2}} \frac{r_{s}}{r} (1 -
e^{-r/\bar{\lambda}}) + C + \frac{r_{s}}{4{r_{0}}^{2}} r - \frac{A}{r} +
\frac{U'({\phi}_{II})}{6} r^{2} \end{equation} The continuity of $\phi$ at the
radius $r_{II}$ involves :  \begin{equation} \label{Phi I} {\phi}_{I} =
\frac{{\bar{\lambda}}^{2}}{2{r_{0}}^{2}} \frac{r_{s}}{r_{II}} (1 -
e^{-r_{II}/\bar{\lambda}}) \end{equation} Further, the above solution involves a
critical radius $r_{c}$ at which the solutions of both regions I and II are
connected.  This critical radius is a solution of the following equation :
\begin{equation} \label{eq critical radius} C + \frac{r_{s}}{4{r_{0}}^{2}} r -
\frac{A}{r} + \frac{U'({\phi}_{II})}{6} r^{2} = 0.  \end{equation} The constant
$C$ is determined from relation (\ref{eq critical radius}) by requiring that $r_{s}$ and $A$ vanish whenever one sets $M$
equal to zero.  One finds $C = - U'({\phi}_{II})/6$ ${r_{c}}^{2}$ and therefore
:  \begin{equation} \label{critical radius} r_{c} = \sqrt{2\zeta} r_{0}
\end{equation} We will neglect throughout the contribution of cosmological
constant type terms to the dynamics of the ordinary matter at the scale of the
solar system since this is known to be very small at present epoch.

\item In region III, the solution is of the form :  \begin{equation} \label{sol
of phi with U III} \phi = {\phi}_{III} + \frac{D}{r} \{ 1 -
\cos{[\frac{2{\pi}}{{\lambda}'} (r - r_{III})]} \} \end{equation} which implies
\begin{equation} \label{grad phi general III} \frac{d{\phi}}{dr} = -
\frac{D}{r^{2}} \{ 1 - \cos{[\frac{2{\pi}}{{\lambda}'} (r - r_{III})]} -
\frac{2{\pi}r}{{\lambda}'} \sin{[\frac{2{\pi}}{{\lambda}'} (r - r_{III})]} \},
\end{equation} where $D$ is a constant of integration and ${\lambda}' =
2\pi/\sqrt{\mid U''({\phi}_{III}) \mid}$ defines a wavelength for the
$\phi$-field in region III.  Hence, $the$ $\phi$-$field$ $would$ $have$ $a$
$damped$ $oscillatory$ $behavior$ $in$ $the$ $regions$ $of$ $space$ $where$
$U''(\phi) < 0$.  \end{itemize}

\section{Einstein equations}

\subsection{The gravitational field sources} The metric tensor $g_{\mu\nu}$ is
solution of the Einstein equations \begin{equation} \label{Einstein eq}
R_{\mu\nu} - \frac{1}{2} R g_{\mu\nu} = \frac{8\pi G}{c^{4}} T_{\mu\nu}.
\end{equation}In the presence of the scalar field $\phi$, its right-hand side
\begin{equation} \label{energy-momentum} T_{\mu\nu} = T^{\circ}_{\mu\nu} +
T^{(\phi)}_{\mu\nu} \end{equation} incorporates the energy-momentum tensor of
the ordinary matter, $T^{\circ}_{\mu\nu}$, and the energy-momentum tensor of the
$\phi$-field itself.  Let us emphasize that the scalar field considered in this
paper is external to gravity (like the electromagnetic field) but obeys the
equivalence principle (unlike the electromagnetic field).

\subsection{The weak fields approximation} Let us denote as
${g^{\circ}_{\mu\nu}}$ (resp.  $g^{\circ\mu\nu}$) the solution of the Einstein
equations for $\phi = 0$ and ${g_{\mu\nu}}$ (resp.  $g^{\mu\nu}$) the components
of the metric tensor in the presence of the $\phi$-field (all greek indices run
over 0, 1, 2, 3 and $x^{0} = ct$) ; $R_{\mu\nu}$ denotes the Ricci tensor, $R =
g^{\mu\nu} R_{\mu\nu}$ is the curvature scalar (Einstein's summation convention
is adopted throughout this paper) and the ${{\Gamma}^{\mu's}_{\alpha\beta}}$ are
the Christoffel symbols.  Hereafter, whenever we assume spherical symmetry :
$x^{1} = r$, $x^{2} = \theta$, $x^{3} = \varphi$ (for the sake of simplicity,
for planar motion $\varphi = \frac{\pi}{2}$ in the following), otherwise the
$x^{i's}$ denote the Cartesian coordinates ($i = 1, 2, 3$).  Einstein equations
rewrite \begin{equation} \label{Einstein eq ext sol} R_{\mu\nu} = \frac{8\pi
G}{c^{4}} [(T^{\circ}_{\mu\nu} - \frac{1}{2} T^{\circ} g_{\mu\nu}) +
(T^{(\phi)}_{\mu\nu} - \frac{1}{2} T^{(\phi)} g_{\mu\nu})], \end{equation} where
$T^{\circ} = g^{\alpha\beta} T^{\circ}_{\alpha\beta}$ is the trace of
$T^{\circ}_{\mu\nu}$ and $T^{(\phi)} = g^{\alpha\beta} T^{(\phi)}_{\alpha\beta}$
is the trace of $T^{(\phi)}_{\mu\nu}$.  In the weak field approximation, one
gets in particular :  $T^{(\phi)}_{00} - \frac{1}{2} T^{(\phi)} g_{00} = -
\kappa (U(\phi) + \int J d\phi)$ and $T^{\circ}_{00} - \frac{1}{2} T^{\circ}
g_{00} = \frac{1}{2} \rho c^{2}$ (weak gravitational field approximation).
Furthermore, one has in the first approximation \begin{equation} \label{Ricci
comp 00} R_{00} = \frac{1}{2} {\nabla}^{2} g_{00}.  \end{equation} So, we may
write :  \begin{equation} \label{g00 first order} {g}_{00} = 1 + 2 \frac{V_{N} -
V_{\phi}}{c^{2}} \end{equation} with \begin{equation} \label{extrapot Vphi}
{{\nabla}^{2}} V_{\phi} = \frac{8\pi G}{c^{2}} \kappa (U(\phi) + \int J d\phi)
\end{equation} ${{\nabla}^{2}} V_{N} = 4\pi G \rho$ and ${g^{\circ}_{00}} = 1 +
2 V_{N}/c^{2}$, where $\rho$ is the density of the ordinary matter.  Derivating
partially equ(\ref{extrapot Vphi}) with respect to $\phi$ then comparing with
equ(\ref{scalar field SR eq}) yields :  \begin{equation} \label{Vphionphi}
\frac{\partial V_{\phi}}{\partial \phi} = (\frac{\partial V_{\phi}}{\partial
\phi})_{r=r_{I}} + \frac{8\pi G}{c^{2}} \kappa \phi, \end{equation} on account
that the derivative $\frac{\partial V_{\phi}}{\partial \phi}$ should be bounded
even when extrapolated at $r = 0$.

\section{Equation of motion}The equation of motion of a test body in the
presence of the scalar field $\phi$ writes in curved spacetime :
\begin{equation} \label{dyn eq} \frac{du^{\mu}}{ds} +
{\Gamma}^{\mu}_{\alpha\beta} u^{\alpha} u^{\beta} = - {\partial}^{\mu} \phi +
\frac{d\phi}{ds} u^{\mu}.  \end{equation}

This means that a force term $F^{\mu} = mc^{2} [{\partial}^{\mu} \phi -
(d\phi/ds) u^{\mu}]$ enters in the right-hand side of the equation of motion of
a test body of mass $m$ in the presence of the $\phi$-field.  The first term of
the right-hand side, ${\partial}^{\mu} \phi$, is analogous to the electric part
of the electromagnetic force whereas the second one $- (d\phi/ds)~u^{\mu}$ is
analogous to the magnetic part.  Both terms are necessary to satisfy the
unitarity of the velocity 4-vector ($u_{\mu} u^{\mu} = 1$, hence $u_{\mu}
(u^{\nu} {\nabla}_{\nu}) u^{\mu} = 0$.  Equation (\ref{dyn eq}) may be derived
from the Lagrangian :  \begin{equation} \label{Lagrangian} L = -
\frac{mc^{2}}{2} e^{-\phi} (g_{\mu\nu} u^{\mu} u^{\nu} + 1).  \end{equation}

\subsection{Motion in weak fields with low velocity} In the weak fields and low
velocity limit, equation (\ref{dyn eq}) simplifies to \begin{equation}
\label{dyn eq approxim} \frac{d^{2}x^{i}} {{dt}^{2}} = - c^{2} {\Gamma}^{i}_{00}
- g^{ii} \frac{\partial{\phi}}{\partial{x}^{i}} c^{2} + \frac{d{\phi}}{dt}
\frac{d{{x}^{i}}}{dt}, \end{equation} Now, $g^{ii} \simeq - 1$ and
${\Gamma}^{i}_{00} \simeq - 1/2~g^{ii}~\partial g_{00}/ \partial x^{i}$.  Hence,
the equation of motion rewrites in vectorial notation :  \begin{equation}
\label{EP dyn eq approxim vectorial} \frac{d^{2}\vec{r}} {{dt}^{2}} = -
\vec{\nabla}V_{N} - f c^{2} \vec{\nabla}{\phi} + \frac{d{\phi}} {dt}
\frac{d\vec{r}} {dt} \end{equation} where we have set $f = \frac{\partial
(V_{\phi}/c^{2})}{\partial \phi} - 1$.  In next section $4.2$, we show that $f$
is positive.

\subsection{Derivation of the long-range acceleration $a_{P}$} The projection of
equation (\ref{EP dyn eq approxim vectorial}) above in plane polar coordinates
$(r,\theta)$ yields, assuming $\mid dr/dt \mid \ll c~\sqrt{f}$, the radial
component of the acceleration vector, \begin{equation} \label{radial acc.}
a_{r} = - \frac{G M} {r^2} - f \frac{d({\phi} c^2)} {dr}.  \end{equation} In the
low velocity limit, the tangential component of the acceleration vector,
$a_{\theta}$, is equal to zero (conservation of the angular momentum).  Clearly,
relation (\ref{radial acc.}) is of the form :  \begin{equation} \label{rad acc.
bis} a_{r} = - (a_{N} + a_{P}) \end{equation} where $a_{N} = \frac{G M} {r^2}$
is the magnitude of the Newtonian radial acceleration and $a_{P}$ is the radial
acceleration induced by the scalar field, \begin{equation} \label{anomalous rad
acc.}  a_{P} = f c^2 \frac{d\phi}{dr}.  \end{equation} Relation (\ref{anomalous
rad acc.}) applies to any region of space out of the central mass.  Besides,
since $d\phi/dr > 0$, $f$ must be positive for $a_{P}$ mimics a missing mass
gravitational field (see section $7$ below).  Hence, it follows that the radial
acceleration induced by the scalar field will be directed towards the central
mass as observed for Pioneer 10/11, Ulysses and Galileo.

\subsubsection{Region I} In region I and for $r \gg \bar{\lambda}$, the scalar
field is expelled out and consequently equation (\ref{EP dyn eq approxim
vectorial}) simplifies to :  \begin{equation} \label{EP dyn eq approxim
vectorial I} \frac{d^{2}\vec{r}} {{dt}^{2}} = - (1 + f_{0}
\frac{{\bar{\lambda}}^{2}}{{r_{0}}^{2}}) \vec{\nabla}V_{N} \end{equation} or
equivalently \begin{equation} \label{EP dyn eq approxim vectorial II}
\frac{d^{2}\vec{r}} {{dt}^{2}} = - G \frac{M + M_{hidden}}{r^{2}} {\vec{u}}_{r},
\end{equation} where $f_{0} = f(0)$, ${\vec{u}}_{r} = \vec{r}/r$ is the radial
unitary vector and $M_{hidden} = f_{0} (\bar{\lambda}/r_{0})^{2} M$ mimics a
hidden mass term (so that the true dynamical mass of the Sun differs from its
luminous mass, $M_{\odot}$, by the amount $f_{0} (\bar{\lambda}/r_{0})^{2}
M_{\odot}$).  It is worth noticing that the kind of missing mass which is
invoked here mimics a spherical distribution of dark matter located within the
Sun rather than a solar halo dark matter.  In this respect, we will distinguish
the hidden mass from dark matter.  Throughout, hidden mass means extra terms
involving the $\phi$-field and that mimic a mass term.  Both hidden mass and
dark matter define the missing mass.\\Furthermore, since the $\phi$-field
respects the equivalence principle, equation (\ref{EP dyn eq approxim vectorial
I}) involves a maximum shift $Z$ on the frequency of a photon given by :
\begin{equation} \label{LPI shift} Z = (1 + f_{0}
\frac{{\bar{\lambda}}^{2}}{{r_{0}}^{2}}) \Delta V_{N}/c^{2}.  \end{equation}
Equation (\ref{LPI shift}) above allows us as yet to put an upper bound on the
possible value of $f_{0}$.  Indeed, analysis of the data from the tests of local
position invariance ("the outcome of any local non-gravitational experiment is
independent of where and when in the universe it is performed") yields a limit
$f_{0} < 2~10^{-4} {r_{0}}^{2}/{\bar{\lambda}}^{2}$ (see C.  M.
Will~\cite{Will}).  The local position invariance is one of the three pieces of
the equivalence principle and, since the $\phi$-field respects the equivalence
principle, this is a crucial test for this field.  As we will see further, the
$\phi$-field passes the current tests.  Indeed, one finds that $f_{0}$ is of the
order $10^{-6}$ (see subsection 5.1.2).  In addition, the study of the possible
effect of dark matter on the motion of the outer planets involves that the
missing mass within the Sun is necessarily less than $10^{-6} M_{\odot}$ (see
Anderson et al.~\cite{Andersond}).  Thence, we may conclude that $\bar{\lambda}
\ll r_{0}$.  Clearly, a $\phi$-field of mass $m_{\phi} \geq
1.8~10^{-17}~eV/c^{2}$ passes all the current tests.

\subsubsection{Region II} Let us neglect for the moment the contribution of the
damped oscillations.  We will also neglect the $\phi$-term in relation
(\ref{Vphionphi}) so that $f \simeq f_{0}$ (i.e., we neglect the anharmonic
terms).  Replacing $\frac{d\phi} {dr}$ by the expression (\ref{grad linear sol})
obtained for region II, relation (\ref{anomalous rad acc.}) yields :
\begin{equation} \label{anomalous rad acc.  bis} a_{P} = a^{\infty}_{P} (1 + 2
\frac{{{r'}_{0}}^{2}}{r^{2}}) \end{equation} where we have set \begin{equation}
\label{rescaling} {r'}_{0} = \sqrt{\zeta} r_{0} \end{equation} and
\begin{equation} \label{asympt.  anomalous rad acc.}  a^{\infty}_{P} =
\frac{f_{0}}{2} \frac{GM}{r_{0}^{2}} \end{equation} turns out to be the
asymptotic radial residual acceleration.  Besides, combining relations
(\ref{critical radius}) and (\ref{rescaling}) above yields \begin{equation}
\label{crit.  rad and rescaling} {r}_{c} = \sqrt{2}~{r'}_{0} \end{equation} and
thence \begin{equation} \label{aPmax} a_{P}(r = r_{c}) = 2 a^{\infty}_{P}
\end{equation} which is also the maximum possible value for $a_{P}$.  Relation
(\ref{crit.  rad and rescaling}) may also be derived by requiring the continuity
of the derivative $d\phi/dr$ at radius $r_{c}$ (neglecting the $\Lambda$ term).

\section{Interpretation of the data} It has been questionned why the Pioneer
effect has gone undetected in the planetary orbits of the Earth and Mars.
Precisely, the Viking ranging data limit any unmodeled radial acceleration
acting on Earth and Mars to no more than $0.1~10^{-8}$ cm/s$^{2}$.  Indeed,
since the Pioneer effect is expected in region II but not in region I, there
must be some critical radius $r_{c}$ which allows one to distinguish between
these two region of space within the solar system.  As region II is defined
about the radius $r_{II}$, one may reasonably consider that $r_{c}$ is of the
order $r_{II}/10$ and accordingly the anomalous acceleration $a_{P}$ should be
negligibly small below the radius $r_{c}/2$.  Indeed, our estimate of $r_{II}$,
given in subsection $5.2$, is in accordance with the estimate of $r_{c}$ given
in subsection $5.1.1$ and both estimations corroborate the fact that the Pioneer
effect is negligibly small below the asteroid belt.  Our scalar field, external
to gravity but which respects the equivalence principle, provides a solution to
both the anomalous radial acceleration observed on the spacecraft and the
absence of a comparable effect on the Earth or Mars.  \\Indeed, it is worth
noticing that all the spacecraft which undergo the Pioneer effect were located
at radii well beyond the orbital radius of Mars when the data were received from
them (the closest spacecraft, Galileo and Ulysses, were in the vicinity of
Jupiter).  \\Moreover, "no magnitude variation of $a_{p}$ with distance was
found, within a sensitivity of $2~10^{-8}$ cm/s$^{2}$ over a range of $40$ to
$60$ AU".  On account of these facts, we conclude that the Pioneer effect is a
distance effect and $a_{p}$ is rather asymptotically constant within the regions
hitherto crossed by the spacecraft.  Above all, the scalar field approach leads
to the same conclusion.

\subsection{Estimate of $a_{P}$ for Pioneer 10/11 using Ulysses data} To start
with, let us recall that no magnitude variation of $a_{P}$ with distance was
found , within a sensitivity of $2~10^{-8}$ cm/s$^{2}$ over a range of $40$ to
$60$ AU (the data analysis of unmodeled accelerations began when Pioneer 10 was
at 20 AU from the Sun).  Thus we may set $a_{P} \approx a^{\infty}_{P}$ for the
Pioneer 10/11.  Since we need to be given at least one point in the curve
$a_{P}$ versus r to be able to determine all the parameters needed, our strategy
will consist to use a piece of information from the Ulysses data (the nearest
point) to compute the Pioneer 10/11 data (the farthest points).  It is worth
noticing that the piece of information considered by itself gives no information
on the magnitude of the long-range acceleration of the spacecraft.  It is this
feature that makes the adopted procedure relevant.  To compute $a^{\infty}_{P}$
we need to estimate first $r_{0}$ and $f_{0}$.

\subsubsection{estimate of $r_{0}$ and $r_{c}$} As one can see, for $r = 2
{r'}_{0}$, relation (\ref{anomalous rad acc.  bis}) implies that $a_{P} =
\frac{3}{2} a^{\infty}_{P}$.  Now, this was observed for Ulysses in its
Jupiter-perihelion cruise out of the plane of the ecliptic (at $5.5$ AU).  This
is also consistent with Galileo data (strongly correlated with the solar
radiation pressure ; correlation coefficient equal to 0.99) if one adopts for
the solar radiation pressure (directed away from the Sun) a bias contribution to
$a_{P}$ equal to $(- 4 \pm 3)~10^{-8}$ cm/s$^{2}$.  Hence, we conclude that
${r'}_{0}$ is approximately equal to half of Jupiter's orbital radius, that is
${r'}_{0} \approx 2.75$ AU and consequently ${r}_{c} \approx 3.9$ AU on account
of relation (\ref{crit.  rad and rescaling}).  Let us assume for the moment
$\zeta$ equal to unity ; this leads to conclude that $r_{0} \approx 2.75$ AU.
Figure 2 shows the shape predicted for the curve $a_{P}/a^{\infty}_{P}$ versus
the radius.  The plot starts from the radius $r_{c} = 3.9$ AU to the radius $r =
60$ AU.

\begin{figure} \centerline{\epsfxsize=12cm \epsfbox{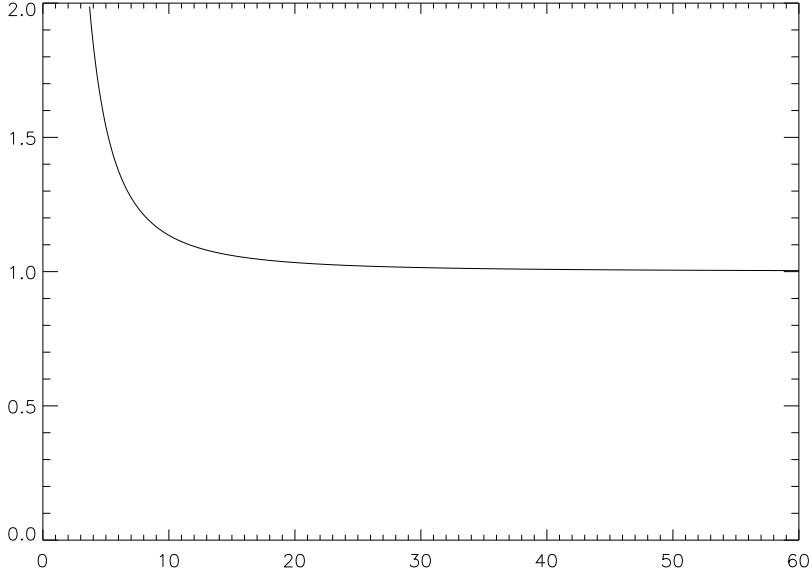}}
\caption{Predicted curve $a_{P}/a^{\infty}_{P}$ versus the radius (in AU) from
$3.9$ AU to $60$ AU.  The curve is asymptotically flat between $40$ AU and $60$
AU (the damped oscillations expected beyond $r_{II}$ have been neglected).}
\end{figure}

\subsubsection{estimate of $f_{0}$ and derivation of the magnitude of
$a^{\infty}_{P}$} In our study on the RC of spiral galaxies, we found that
$f_{0}$ is of the order $\frac{v_{max}^{2}}{c^{2}}$, where $v_{max}$ denotes the
maximum rotational velocity.  This seems to be a general order of magnitude for
this parameter.  So, in what follows, we derive an estimation of $f_{0}$ using
the relation $f_{0} \simeq \frac{v_{max}^{2}}{c^{2}}$, where $v_{max}$ is a
maximum velocity to be determined for the solar system.  In the case of interest
in this paper, the $\phi$-field under consideration though external to gravity
is generated from the Sun (or any other star we would have considered).
Therefore, it seems natural that $v_{max}$ should be a typical velocity that is
related to the matter components of this star and not a peculiar orbital
velocity.  A suitable value for $v_{max}^{2}$ (see Ciufolini~\cite{Ciufolini}),
perhaps the best for it involves thermodynamics parameters solely, is given by
the ratio $P_{c}/{\rho}_{c}$ (assuming the perfect gas), where $P_{c}$ and
${\rho}_{c}$ denote respectively the central pressure and mass density of the
star under consideration.  Taking the value of $T_{c}$ given by solar models
(Stix~\cite{Stix}, Brun et al.~\cite{Brun}), the expression $v_{max}^{2} =
\frac{P_{c}}{{\rho}_{c}}$ gives for the Sun :  $f_{0} = (1.72 \pm 0.04)~10^{-6}$
and $a^{\infty}_{P} = (6.8 \pm 0.2)~10^{-8}$ cm/s$^{2}$, in good agreement with
the recent results which give $a^{\infty}_{P} = (7.29 \pm 0.17)~10^{-8}$
cm/s$^{2}$ as the most accurate measure of the anomalous acceleration of Pioneer
10 (Turyshev et al.~\cite{Turyshev}).  Further, the value computed for
$a^{\infty}_{P}$ may be corrected to $a^{\infty}_{P} = (7.23 \pm 0.2)~10^{-8}$
cm/s$^{2}$ by identifying $\lambda$ with $r_{0}$ (see subsection $5.2$ below) :
hence, $\zeta \simeq 1.07$.

\subsection{Damped oscillations and vanishing of $a_{P}$} Figure 1 of the paper
of Anderson et al.  shows an almost harmonic oscillation of $a_{P}$ (nothing is
said about this by the authors themselves though) for Pioneer 10 which starts at
the radius $r_{II} = 56.7 \pm 0.8$ AU with an amplitude $a_{Pm}$ of the order
$\frac{1}{4}~10^{-8}$ cm s$^{-2}$ (this is derived by comparison with the
uncertainty on $a_{P}$ for Pioneer 10) and a wavelength $\lambda = 2.7 \pm 0.2$
AU the value of which turns out to be quite identical to that of $r_{0}$ (let us
notice by passing that $r_{II}/\lambda$ is an integer (= $21$)).  With these
observational data, relation (\ref{derivative of phi II-III bis}) involves, for
$\zeta \simeq 1$ and $B = A/16$, $a_{Pm} \simeq 0.26~10^{-8}$ cm s$^{-2}$
between $56$ AU and $60$ AU as can be seen in figure 3.

\begin{figure} \centerline{\epsfxsize=12cm \epsfbox{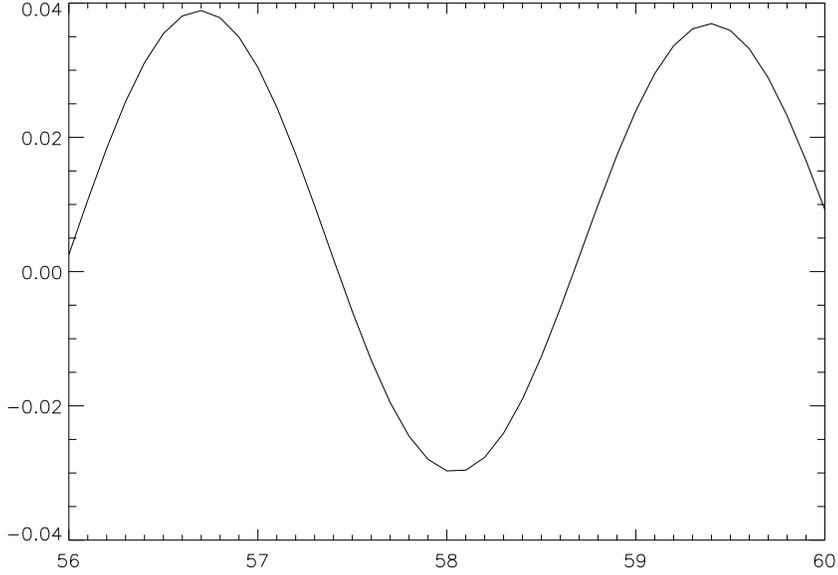}}
\caption{Predicted curve $(a_{P} - a^{\infty}_{P})/a^{\infty}_{P}$ versus $r$
(in AU), for region II with parameters $r_{II} = 56.7$ AU, $\lambda = 2.7$ AU,
$\zeta = 1$ and $B = A/16$.}  \end{figure}

Furthermore, the calculations carried out in subsection $2$ lead to predict the
decline of $a_{P}$ in the form of damped oscillations beyond $r = r_{III}$.
Hence, since $r_{III} > \sqrt{3}~r_{II}$, we may confidently expect the decline
of $a_{P}$ to occur only beyond $r = 96.8$ AU.  Let us emphasize that the
spatial periodicity $\lambda$ involves a temporal periodicity $T_{P} =
\lambda/v_{P}$, where $v_{P}$ is the speed of the spacecraft.  Hence, the
periodicity of one year found for Pioneer 10 has nothing to do with the orbital
periodicity of the Earth.  Indeed, the coincidence just comes from the fact that
$v_{P} = 2.66$ AU/yr for Pioneer 10 (at least since 1987).  As a consequence,
$T_{P}$ should be greater than one year for Pioneer 11 since Pioneer 10 is
faster mooving than Pioneer 11.

\section{Derivation of the steady frequency drift using the equivalence
principle} Since, the $\phi$-field obeys the equivalence principle, the steady
frequency drift may be explained in another way than the Doppler effect thanks
to this principle (cf.  Misner et al~\cite{Misner}).  Actually, the steady
frequency drift and the corresponding "clock acceleration" $- a_{t} = -
2.8~10^{-18} s/s^{2}$ shown by the Compact High Accuracy Satellite Motion Progam
analysis of Pioneer 10 data may also be interpreted as the analogous of the
gravitational redshift linked to the extra potential term $V_{P} = \int a_{P}
dr$ associated to the scalar field.  Indeed, the frequency drift
$\frac{d\Delta\nu}{dt}$ as well as the clock acceleration $- a_{t} =
\frac{d(\frac{\Delta\nu}{\nu})}{dt}$ follow from the relation
$\frac{\Delta\nu}{\nu} = \frac{V_{P}(r_{\oplus}) - V_{P}(r)}{c^{2}} = -
\frac{1}{c^{2}} \int_{{r}_{\oplus}}^{{r}_{\oplus} + ct} a_{P} dr$, where
$r_{\oplus}$ denotes the orbital radius of the Earth and $r = r_{\oplus} + ct$
(one way, as considered by the authors) is the distance of the spacecraft from
the Earth.  Therefore, on account that $dr = c~dt$ for the photons (one way),
one obtains the observed relation $a_{P} = a_{t} c$.  In this way, the identity
$a_{P} = a_{t} c$ seems more natural since, in this approach, it is indeed the
photons received on Earth from the spacecraft that are concerned instead of the
spacecraft themselves.

\section{Conclusion} In this paper, we have presented a possible explanation of
the "Pioneer effect", without being in conflict with the Viking data or the
planetary ephemeris.  This is based on a possible interaction of the spacecraft
with a long-range scalar field, $\phi$, which respects the equivalence
principle.  Like any other form of matter-energy, the $\phi$-field is a
gravitational source through its energy-momentum tensor.  Conversely, its source
is the Newtonian potential of the ordinary matter (in this case, the Sun).  The
calculations were performed in the weak fields approximation, with $U$ a quartic
self-interaction potential.  They gave, near the spacecraft, a residual radial
acceleration directed towards the Sun, with a magnitude $a_{P}$ asymptotically
constant (in region II).  Both $a_{P}$ and the corresponding clock acceleration,
$a_{t}$, computed from our formulas are in fairly good agreement with the
observed values.  Moreover, a scalar field of mass $m_{\phi} \geq
1.8~10^{-17}~eV/c^{2}$ will be expelled from region I in a way quite analogous
to the Meisner effect in a superconducting medium.  This limits $a_{P}$ to no
more than $0.1~10^{-8}$ cm/s$^{2}$, from the radius of the Sun to ${r}_{c} =
3.9$ AU.  It is also found that the $a_{P}$ term should be accompanied with
damped oscillations in the intermediary region between region II and region III.
We also predict beyond ${r}_{III} = 97$ AU (that is, about the year 2009 or so,
for Pioneer 10) the vanishing of $a_{P}$ in the form of damped
oscillations.\\Furthermore, the scalar field theory, as developped in this
paper, also gives good fits for the rotational curves of spiral galaxies as
shown in a previous study.  Moreover, the same field acts at the cosmological
scales like a cosmological constant.  Preliminary estimations (Mbelek and
Lachi\`eze-Rey, in preparation) from the dynamics of the external region of the
dwarf galaxy DDO 154, actually the sole galaxy for which the edge of the mass
distribution has been reached (see Carignan \& Purton~\cite{Carignan}) led to a
value ${\Omega}_{\Lambda} = 0.43$~(H$_{0}$/100~km~s$^{-1}$~Mpc$^{-1}$)$^{-2}$ in
fairly good agreement with the value ${\Omega}_{\Lambda} = 0.7$ deduced from the
Hubble diagram of the high-redshift type Ia supernovae (Perlmutter et
al.~\cite{Perlmutter}, Schmidt et al.~\cite{Schmidt}, Riess et al.~\cite{Riess},
Garnavich et al.~\cite{Garnavich}) for H$_{0}$ about $75$ km s$^{-1}$/Mpc.

\end{document}